\pdfoutput=1

\documentclass[%
 aip,
 amsmath,amssymb,
 reprint,%
]{revtex4-1}

\usepackage{graphicx}% Include figure files
\usepackage{dcolumn}% Align table columns on decimal point
\usepackage{bm}% bold math
\usepackage{amsmath}
\usepackage{amssymb}
\usepackage[utf8]{inputenc} % allow utf-8 input
\usepackage[T1]{fontenc}    % use 8-bit T1 fonts
\usepackage{hyperref}       % hyperlinks
\usepackage{url}            % simple URL typesetting
\usepackage{booktabs}       % professional-quality tables
\usepackage{amsfonts}       % blackboard math symbols
\usepackage{nicefrac}       % compact symbols for 1/2, etc.
\usepackage{microtype}      % microtypography
\usepackage{graphicx,xcolor}
\usepackage{mathptmx}
\graphicspath{{images/}}
\usepackage{amsmath,tabularx}

\DeclareMathOperator{\EX}{\mathbb{E}}% expected value
\usepackage{esvect}
\usepackage{wrapfig}
\DeclareMathOperator{\E}{\mathbb{E}}

\begin{document}
\preprint{AIP/123-QED}

\title{Temperature-transferable coarse-graining of ionic liquids with dual graph convolutional neural networks}

\author{Jurgis Ruza}
\affiliation{Materials Science and Engineering, \'Ecole Polytechnique F\'ed\'erale de Lausanne, Lausanne, Switzerland}
\affiliation{Department of Materials Science and Engineering, Massachusetts Institute of Technology, Cambridge, Massachusetts}
\author{Wujie Wang}
\author{Daniel Schwalbe-Koda}
\affiliation{Department of Materials Science and Engineering, Massachusetts Institute of Technology, Cambridge, Massachusetts}
\author{Simon Axelrod}
\affiliation{Department of Materials Science and Engineering, Massachusetts Institute of Technology, Cambridge, Massachusetts}
\affiliation{Department of Chemistry and Chemical Biology, Harvard University, Cambridge, Massachusetts}
\author{William H. Harris}
\author{Rafael G\'omez-Bombarelli}
\email{rafagb@mit.edu}
\affiliation{Department of Materials Science and Engineering, Massachusetts Institute of Technology, Cambridge, Massachusetts}

\begin{abstract}
Computer simulations can provide mechanistic insight into ionic liquids (ILs) and predict the properties of experimentally unrealized ion combinations. However, ILs suffer from a particularly large disparity in the time scales of atomistic and ensemble motion. Coarse-grained models are therefore used in place of costly all-atom simulations, accessing longer time scales and larger systems. Nevertheless, constructing the many-body potential of mean force that defines the structure and dynamics of a coarse-grained system can be complicated and computationally intensive. Machine learning shows great promise for the linked challenges of dimensionality reduction and learning the potential of mean force. To improve the coarse-graining of ILs, we present a neural network model trained on all-atom classical molecular dynamics simulations. The potential of mean force is expressed as two jointly-trained neural network interatomic potentials that learn the coupled short-range and many-body long range molecular interactions. These interatomic potentials treat temperature as an explicit input variable to capture its influence on the potential of mean force. The model reproduces structural quantities with high fidelity, outperforms the temperature-independent baseline at capturing dynamics, generalizes to unseen temperatures, and incurs low simulation cost.
\end{abstract}

\maketitle

\section{Introduction}

Ionic liquids (ILs) are room temperature liquids composed of a molecular cation and anion. ILs have a wide variety of applications, including ion transport in batteries and catalysis \cite{Dupont2002, Lin2015}. Because of the wide chemical diversity of molecular anions and cations, and their many possible combinations, a vast design space for ILs exists. Molecular dynamics simulations can characterize the structure and dynamics of ILs at the nanoscale, and, to some degree, predict their properties before experimental realization. However, due to the complex intermolecular forces in ILs, classical force fields need to be fine-tuned to recover the correct kinetics and structure\cite{CanongiaLopes2006, CanongiaLopes2004ModelingField}. The properties of ILs are governed by a variety of interactive forces, including weak and isotropic forces such as van der Waals, solvophobic and dispersion forces \cite{Greaves2013, Izgorodina2014}, strong Coulombic forces, anisotropic hydrogen bonding \cite{Fumino2014}, halogen bonding \cite{Mukai2013}, and magnetic dipole, electron pair and dipole-dipole interactions \cite{Kashyap2010}. 
Further complicating simulation is the sluggish motion of the ensemble. This leads to a large gap in the time scales of atomistic and ensemble motion, limiting the size of the molecular system and the time scales that can be accessed through simulation. 

Coarse-graining (CG) can increase accessible simulation sizes by reducing the dimensionality of the full atomistic model to a pseudoparticle representation.\cite{Izvekov2005, Noid2008} 
A coarse-grained model can be described by a potential energy surface, the many-body potential of mean force (PMF). The PMF can be seen as the configurational free energy in a reduced phase space and is determined by the all-atom potential and the mapping operator from a full to a coarse-grained resolution.\cite{Noid2013}

However, constructing an accurate many-body PMF that can faithfully capture the structure, thermodynamics and kinetics of the coarse-grained system is not an easy task. This is because reducing the representation and averaging out the fast atomic motions introduces complex many-body correlations in the PMF. Furthermore, the PMF is conditioned on the choice of mapping from the all-atom to the compressed model and on the temperature of the reference data. Typically, these mappings are chosen based on chemical intuition and trial-and-error, although there exist some methods for automatic and statistical mapping algorithms.\cite{Chen2006AnPolyethylene, Lombardi2016CG2AA:Structures, Wang2019}

CG-IL models are particularly challenging because of the different transport properties of the two components, which further complicates the modeling of the many-body PMF.\cite{Kowsari2008, Umecky2005, Sangoro2008ElectricalLiquid} Many authors have developed CG-IL models, and while each has achieved some success, none has been completely satisfactory. For example, Refs.\cite{Roy2010, Wang2009, Merlet2012, Wang2013} studied the 1-butyl-3-methylimidazolium cation with different anions. In one case, idealized models recovered the structural and energetic properties, but their transport properties were unreasonably slow. \cite{Roy2010} The model was improved upon by Merlet \textit{et al}.\cite{Merlet2012}, and the transport properties were recovered to a higher degree, but with a loss of structural properties.\cite{Merlet2012} Wang \textit{et al.} \cite{Wang2009} described an effective force coarse-graining applied to an IL system and recovered structural (RDF) properties over a range of temperatures, but they did not investigate dynamics. Ref.\cite{Wang2013} described Newton Inversion and Iterative Boltzmann Inversion methods for coarse-graining ILs, and the thermodynamic properties were well recovered. However, the model underperformed with respect to self-diffusivity, both in proximity to experimental data and in temperature scaling.\cite{Wang2013} Moreover, ILs exist as liquids over a wide range of temperatures, and many of their desired applications involve temperature swings, but none of the CG approaches mentioned above are temperature transferable. Although energy renormalization has been applied \cite{Xia2017} for temperature-transferable CG focused on recovering structure of ILs, work on temperature transferability is lacking.

Each of the methods above achieved success in many, but not all, aspects of coarse graining. Moreover, each method required user expertise to create CG models that were not transferable to other ionic liquids. An alternative approach is to use machine learning (ML). ML approaches can achieve dimensionality reduction in a variety of domains\cite{Gomez-Bombarelli2018,Wang2019} and learn complex functions such as the CG many-body PMF.\cite{Wang2019,Nuske2019,Wang2019a,Wang2020} Neural coarse graining has been shown to recover ensemble quantities of liquid-phase alkanes with high accuracy, and without expert user input.\cite{Wang2019} Here we expand on this bottom-up approach to learn the encoding function and the associated PMF for an 1-butyl-3-methylimidazolium tetrafluoroborate system, with two key extensions. First, we condition the PMF on the temperature, achieving temperature-transferable dynamics at temperatures between two extremes. Second, in order to capture the complex many-body intermolecular forces acting on CG ILs, we utilize two separate neural network models to learn the coupled intra- and inter-molecular PMF. These two key features allow coarse-graining more complex condensed phase systems like ILs and offer a reliable, automated alternative to expert-based IL coarse graining.

\begin{figure}[htb]
    \centering
    \includegraphics[width=.42\textwidth]{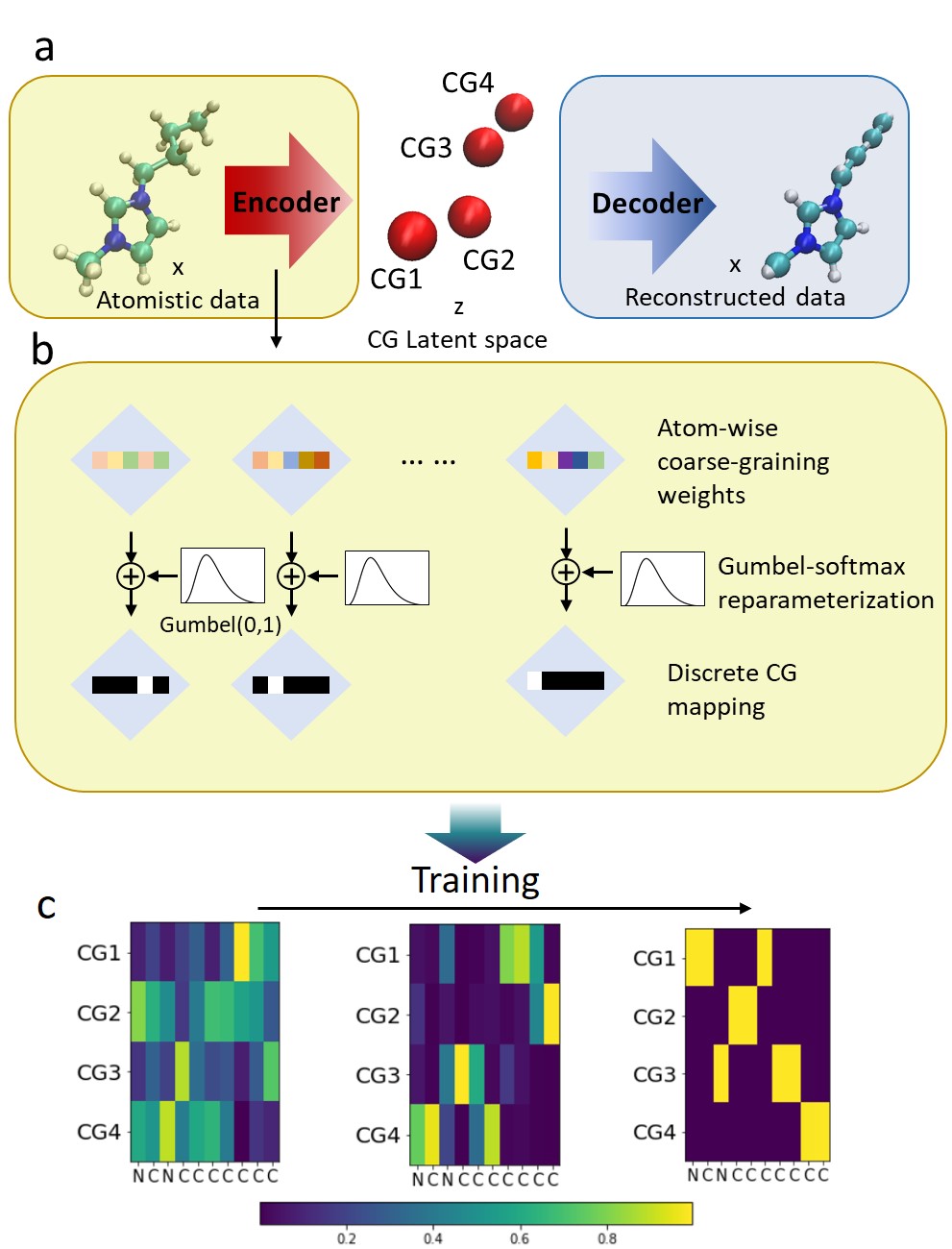}
    \caption{Model framework of the coarse-grained auto-encoder. \textbf{a} - The encoder-decoder architecture is optimized by minimizing the sum of the reconstruction and force regularization losses (Equation \ref{eq:loss}). \textbf{b} - To achieve discrete mapping, the computational graph is optimized using the Gubmel-softmax reparametrization, with weights randomly initiated as atom-wise vectors. \textbf{c} - Example of the training scheme with the convergence to an atom-wise mapping, the x-axis represents each heavy atom in the cation and the y-axis the contribution to each bead.}
    \label{autograin}
\end{figure}{}

\section{Methods}

\subsection{Auto-encoder coarse-graining}

The general schematic of auto-encoder coarse graining is shown in Fig. \ref{autograin}. The model is an autoencoder, consisting of an encoder that maps the atomistic data to a coarse-grained latent representation, and a decoder that maps the latent representation back to the atomistic representation. To train the model we minimize the autoencoder loss, $L_{AE}$, which was introduced in Ref. \cite{Wang2019} as:
\begin{equation}
    L_{AE} = \frac{1}{N} \EX_{x\sim P(x)}[(D(E(x,g,\tau)) - x)^2+\rho F_{inst}(E(x))^2].
    \label{eq:loss}
\end{equation}{}

The first term in Eq. \ref{eq:loss} is a reconstruction loss, where $D$ is the decoder mapping operator from the coarse-grained variables to the all atom representation, $E$ is the encoder mapping operator from the all atom representation to the coarse-grained beads, $x$ is the observed atomistic samples from all-atom MD snapshots (atomic number and cartesian coordinates of every atom in the all-atom MD snapshot). $N$ is the number of coarse grained beads. $g$ is sampled from a Gumbel distribution and is used for learning the categorical, discrete coarse-grained weights for each atom using Gumbel-softmax reparametrization.\cite{jang2016categorical} $\tau$ is a temperature-like variable that decreases along training to achieve discrete encoding in the coarse-grained mapping $E$. The second term is an instantaneous force regularization, minimizing of it produces less noisy mapped forces ($F_{inst}$) of a therefore smoother coarse-grained energy landscape. $F_{inst}$ is the instantaneous mapped force acting on the CG beads, and is calculated from the encoding function and the atomic forces on a given all-atom MD frame, $F_{inst}(z) = - \mathbf{b} \nabla V(x)$. The reconstruction loss statistically minimizes the loss of geometrical information that occurs when going through the latent space, and the force regularization is used to minimize the fluctuations in the mean force of the encoded space. By choosing the CG mapping that minimizes force fluctuations, a smoother energy landscape is achieved and more intuitive mappings are obtained. A weight parameter $\rho$ is introduced as a heuristic to adjust the balance between the two terms.

\subsection{Temperature-dependent graph neural network potentials}

Once a mapping $E$ was chosen, a matching PMF was fitted through a graph convolutional neural network based on the SchNet model. \cite{Schutt2018} In the SchNet model the geometry is transformed into a set of atomic fingerprints (the ``convolution'' component), and a fully connected network transforms the fingerprints into an output (the ``readout'' component). The convolution blocks consist of a message step and update step to systematically gather information from neighboring atoms. A neighbor is defined as an atom within a fixed distance cutoff (discussed further below). Defining $v$ as index for each atom and its neighbors as $N(v)$, the $t^{\mathrm{th}}$ graph convolution updates the atomic feature vectors $h_v$ by aggregating ``messages'' from their connected atoms $v$ and their edge features $e_{uv}$, through \cite{Gilmer2017}:
\begin{equation} \label{eq:gcn}
	h_v^{t} = h_v^{t-1} + \sum_{u \in N(v)} \mathrm{message}^{t}(h_u , e_{uv})
\end{equation}
The initial feature vector $h^0$ is generated from a learnable embedding of atomic numbers. By performing the convolution operations several times, a many body correlation function can be parametrized to represent the many body potential of mean force of the coarse-grained ionic liquid complex. In the case of SchNet, the update function is simply a sum over the atomic embeddings. The message function is parameterized by

\begin{equation} \label{eq:message}
 \mathrm{message}^{t}(e_{uv}, h_v) = MLP_3( MLP_1(e_{uv}) \circ  MLP_1(h_v)),
\end{equation}
where MLP denotes a multi-layer perceptron and $\circ$ is the concatenation operation. The readout consists of a sum over MLPs acting on each of the $N$ atomic embedding at the final ($T^{\mathrm{th}}$) convolution: \begin{align}
    \mathrm{Energy} = \sum_{v}^{N} MLP(h_v^T). \label{eq:schnet_readout}
\end{align}
Further details of the original model can be found in Ref. \cite{Schutt2018}.

\begin{figure}[htb]
    \centering
    \includegraphics[width=.40\textwidth]{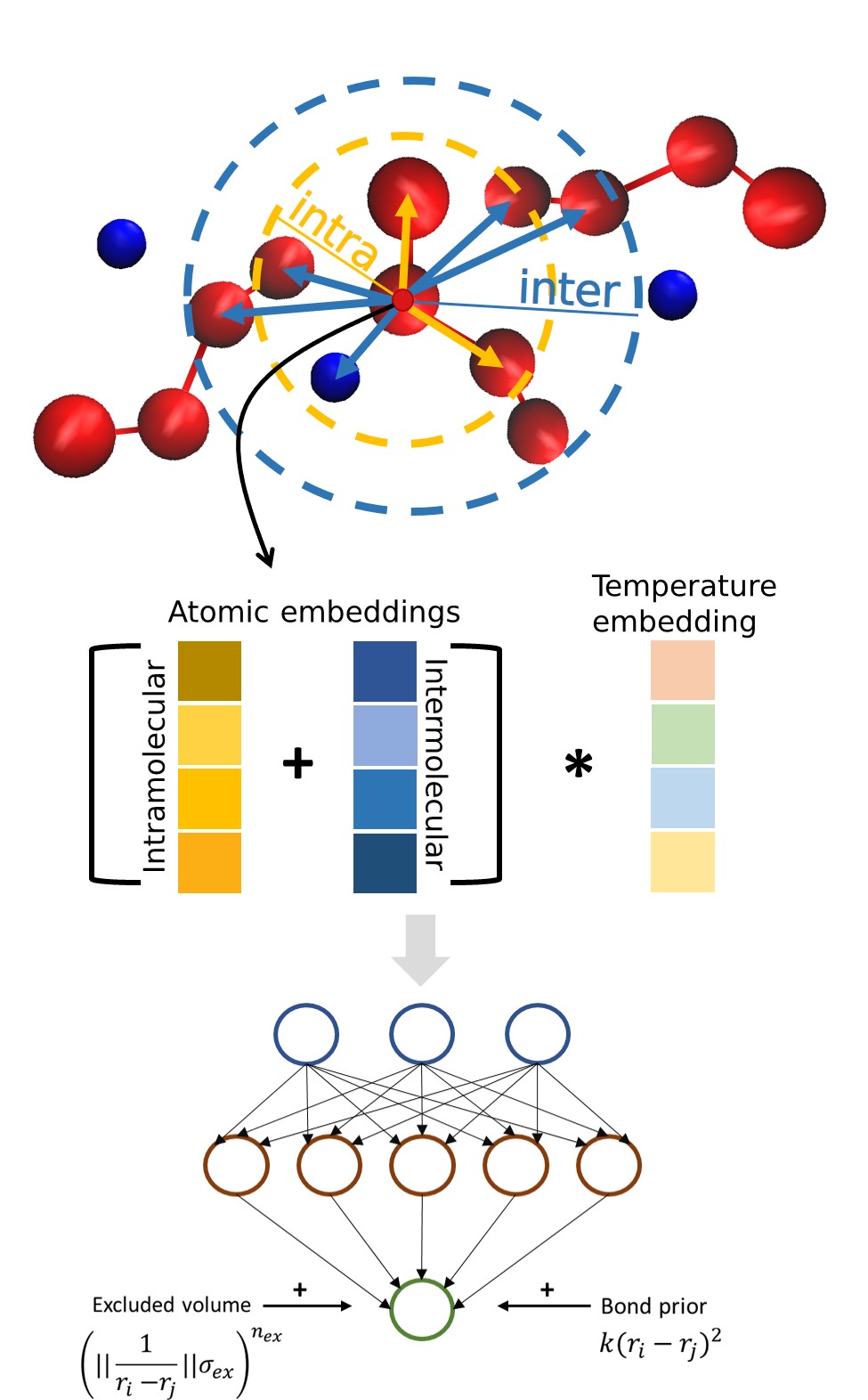}
    \caption{Overview of the neural network architecture, which is based on the SchNet model \cite{Schutt2018}. Instead of having a single interaction block, two are used to model the inter- and intra-molecular interactions of the coarse-grained atoms. Temperature learning and classical priors in the forms of excluded volume and bond length are also used.}
    \label{hybridgraph}
\end{figure}{}

Figure \ref{hybridgraph} shows the neural network architecture for fitting a coarse-grained potential. The model is based on the SchNet architecture, but uses graph convolutions performed at two scales through two separate networks [as defined in Eq. (\ref{eq:gcn})]. Specifically, an intramolecular embedding ($h_{v,\; intra}$) is calculated by performing graph convolutions only among atoms belonging to the same molecule and within an intramolecular distance threshold ($d_{intra}$). A second intermolecular embedding ($h_{v, \; inter}$) is calculated by  performing graph convolutions among atoms non belonging to the same molecule, and within an intermolecular distance threshold ($d_{inter}$). $d_{intra}$ and $d_{inter}$ are hyperparameters chosen through hyperparameter optimization, typically with $d_{inter} > d_{intra}$. The parallel graph convolution passes create two embeddings for each CG bead. These embeddings are summed and the single resulting fingerprint then fed into a multi-layer perceptron for atomic energy parametrization. The per-atom energies are summed into a total energy, and its derivative fitted through stochastic force matching. The purpose of this restrictive design was to differentiate the atomic environment after coarse-graining: the convolution procedure can differentiate its CG bead neighbor by inter-molecular interactions and intra-molecular interactions. This is necessary because intra- and inter-molecular distances in CG systems are closer than in all-atom simulations, but have distinct and complicated functional forms. For example, covalently linked beads in the cation interact through different mechanisms than non-bonded ion pairing regardless of distance. Even point charge interactions become complex anisotropic functions in the PMF landscape, yet occur over distances comparable to intramolecular interactions.

A classical bond prior with fixed bond lengths between the cation pseudoparticles to achieve better learning of the intramolecular interactions,

\begin{equation}\label{eq:bondprior}
    F_{bond} = k (\textbf{r}_i-\textbf{r}_j)^2,
\end{equation}

\noindent where $k$ is a spring constant hyperparameter, and $r_i$, $r_j$ are the positions of neighboring particles. Furthermore, an excluded volume prior was added to obtain better intermolecular interactions:

\begin{equation}\label{eq:excluded}
    F_{ex} = \left(||\frac{1}{\textbf{r}_i-\textbf{r}_j}|| \sigma_{ex}\right)^{n_{ex}},
\end{equation}

\noindent with $\sigma_{ex}$ and $n_{ex}$ hyperparameters related to the excluded volume.

We further included temperature as a neural network input to parameterize a temperature-dependent potential of mean force. We considered that temperature induces fluctuations and biases in the mean forces and thereby affects structural correlation. To this end, we transformed the inverted temperature $T$ into an embedding to be combined with the convolution operations, thereby modifying the structural representation. In particular, we generated a thermodynamic embedding $\gamma$ according to
\begin{equation} \label{eq:temp}
	\gamma = \tilde{M}T^{-1},
\end{equation}
where $\tilde{M}$ is a linear matrix of size $1 \times N_{basis}$ that maps the temperature dependence into the same size as the atomic embedding and $T$ is the thermostat temperature of the all-atom NPT simulation.
After the convolution layers, the intramolecular embedding was summed with the intermolecular embedding; the dot product of the atomic and thermodynamic embeddings was taken. A shallow, 2-layer MLP was used as readout layer to map individual embeddings into atomic energies as in Eq. (\ref{eq:schnet_readout}). 

\begin{equation} \label{eq:readout}
    \mathrm{Energy} = \sum_{v}^N MLP( (h_{v, \; intra}^{T}  + h_{v,\; inter}^{T} ) \cdot \gamma )+ F_{ex} + F_{bond}
\end{equation}

\subsection{Stochastic force matching}

The force-matching approach for parameterizing force fields was proposed in Ref. \cite{Izvekov2005} to reproduce structural correlation functions. Given an atomistic potential energy function $V(x)$ with the partition function $Z$, the probabilistic distribution of atomistic configurations is:
\begin{equation} \label{eq:1}
P(x) = \frac{1}{Z} e^{- \beta V(x)}.
\end{equation}
The distribution function of coarse-grained variables $P_{CG}(z)$ and the corresponding many-body potential of mean force $A(z)$ is the log-likelihood of the distribution of coarse-grained variables. It is a free energy obtained from marginalizing over micro-states that are mapped to the coarse-grained coordinates,
\begin{equation} \label{eq:3}
A(z) = -\frac{1}{\beta} \ln P_{CG}(z).
\end{equation}
The mean force of the coarse-grained variables is the average of instantaneous forces conditioned on the mapping $\mathbf{b} x = z$ \cite{Kalligiannaki2015TheSystems, Darve2006}:
\begin{equation} \label{eq:4}
-\frac{dA}{dz} = F(z)  = \langle -\mathbf{b} \nabla V(x) \rangle_{E(x) = z}
\end{equation}
where $\mathbf{b}$ corresponds to the mapping assignment from the all-atom to the coarse-grained representation. Force matching requires the minimization of the loss function:

\begin{equation} \label{eq:FM}
\min_{\theta}L = \min_{\theta} \E[(F(z) + \nabla_z V_{CG}(z))^2]
\end{equation}
where $\theta$ are the parameters in $V_{CG}$ and $\nabla V_{CG}$ represents the ``coarse-grained forces'' which can be obtained from automatic differentiation. As proposed in Zhang \textit{et al.}, if $\E_z[F_{inst}] = F(z)$ is assumed, the loss function can be reformulated as the following minimization target:
\begin{equation} \label{eq:10}
\min_{\theta} L_{inst} = \min_{\theta} \E[F_{inst}(z)+ \nabla V_{CG} (z))^2].
\end{equation}
This minimization target has also been used in several works with stochastic gradient descent. \cite{Wang2019, Wang2019a} Instead of matching mean forces that need to be obtained from constrained dynamics, the model minimizes $L_{inst}$ with respect to $V_{CG}(z)$ and $E(x)$. This target provides a data-driven target for optimizing coarse-grained force fields.

\section{Results}

\begin{figure}[t!]
    \centering
    \includegraphics[width=0.4\textwidth]{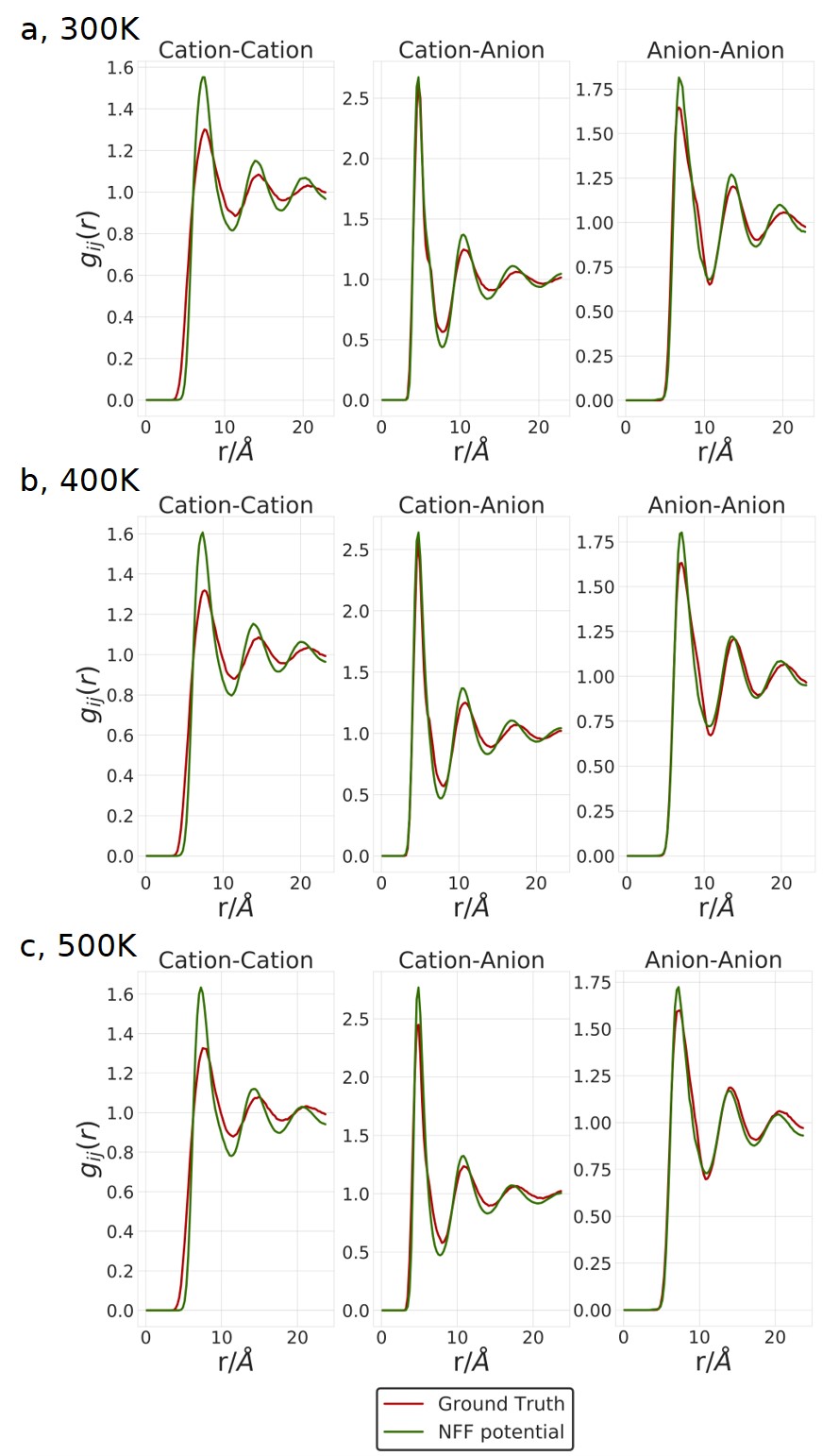}
    \caption{Radial distribution functions of the ionic liquid system with 300 cation and anion molecules. GT denotes the ground truth and T-NFF is the temperature transferable neural force field. Results are shown at temperatures of a) 300 K, b) 400 K (out of sample), c) 500 K.}
    \label{RDF}
\end{figure}{}

The model was trained on molecular dynamics simulations performed with the Large-scale Atomic/Molecular Massively Parallel Simulator (LAMMPS) engine \cite{Plimpton1993}, with a  publicly available custom force field for ionic liquids \cite{CanongiaLopes2006, CanongiaLopes2004ModelingField}. The simulation box was created with a size of 300 ion pairs with PACKMOL \cite{Martinez2009}. A 1 ns NPT simulation with a Nos\'e-Hoover thermostat and barostat was run to equilibrate the box followed by a 10 ns simulation performed at different temperatures, each repeated 3 times with different randomly initiated starting velocities to obtain larger space sampling. A subsample of the dataset was used for training: out of the 30,000 timesteps, 6,000 were evenly sampled and used further. The model was then trained on data from 300, 350, 450 and 500 K simulations, with hyperparameters chosen using Bayesian hyperparameter optimization as detailed in the Supplementary Information section. New CG molecular dynamics simulations with the trained models were run at 300-500 K in 50 K increments, thus including experiments at 400 K, outside of the training data. The new CG simulations were run at NVT ensamble with the Nose-Hoover thermostat at a timestep of 1fs. The structure and dynamics of CG simulations was then compared with the all-atom ground truth.

In keeping with the convention in the field, we utilized 1 bead to represent the anion. The choice of number and mapping of beads to represent the cation is less settled. Here, we  tested two representations of 1-butyl-3-methylimidazolium through either 2 or 4 pseudoparticles (CG3 and CG5 respectively, when counting anion bead).

\begin{figure}[htb]
    \centering
    \includegraphics[width=.48\textwidth]{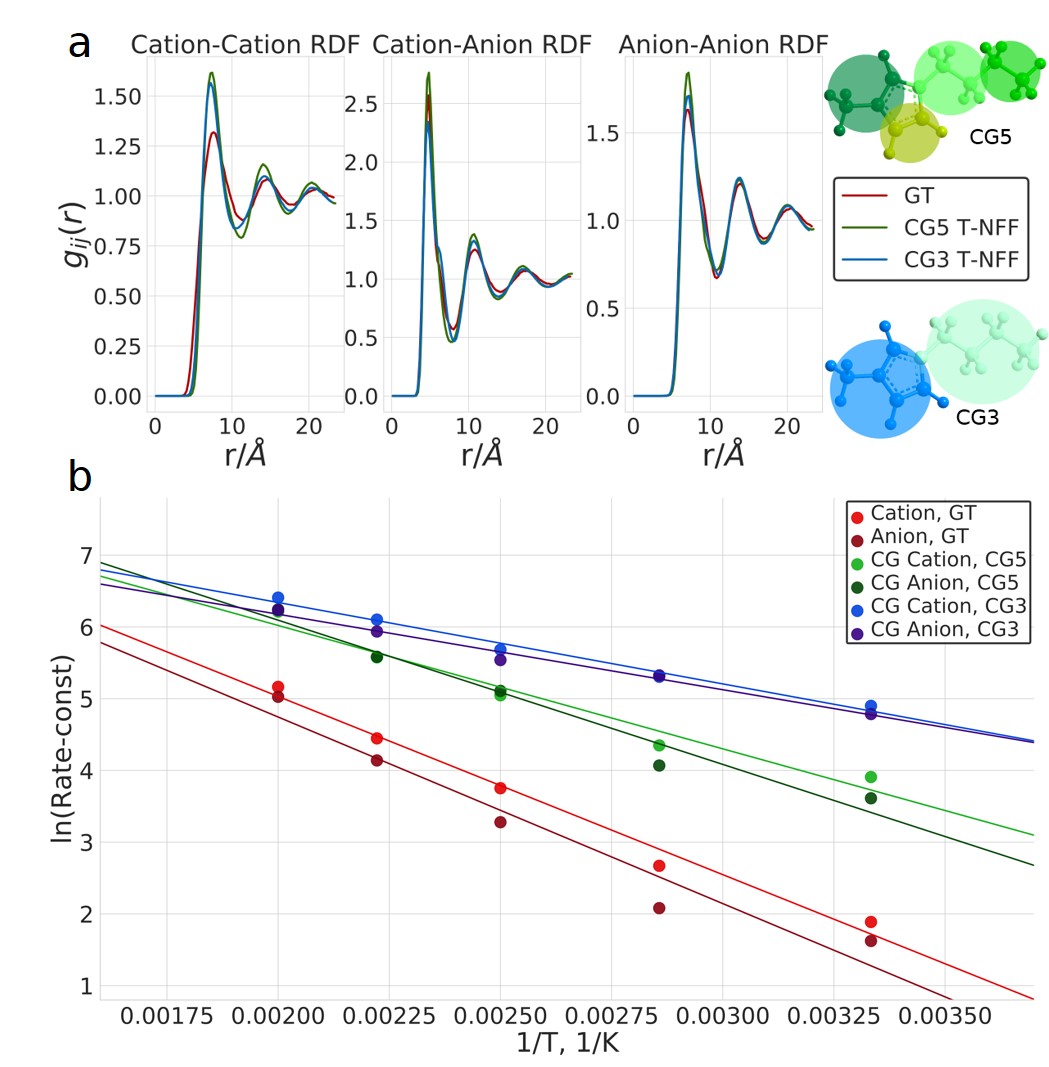}
    \caption{Comparison of models trained on a 3 bead representation (2 cation and 1 anion, CG3) and on a 5 bead representation (4 cation and 1 anion, CG5). Both models use the temperature transferable model (T-NFF). a) RDF of the ground truth, temperature transferable model with 3 and 5 bead representations, ran at 400 K. b) Arrhenius plot comparison of the models.}
    \label{compare_3cg}
\end{figure}

Figure \ref{autograin} \textbf{a} shows the mapping and average reconstruction of 1-butyl-3-methylimidazolium. As expected, the 4-bead model led to good reconstruction, since it captures the orientation of the alkyl chain. The assignment learned by the auto-encoder structure agrees with prior intuition: one bead for the side methyl group and two ring atoms, one bead for the two following carbon atoms in the ring, one bead for the other imidazolium nitrogen and two connected linear carbons (Fig. \ref{compare_3cg}), and a final bead for the terminal ethyl group. Because the model only learns to reconstruct average positions, rotationally equivalent atoms were typically reconstructed near their center of gyration.

Figure \ref{RDF} shows the radial distribution functions (RDF) of the CG5 model (4-bead cation plus tetrafluoroborate anion). The ground truth reference data is the result of applying the CG mapping to held out all-atom frames. The RDFs are in close agreement with the ground truth data. Both the cation-anion and anion-anion interactions have been reproduced to a high degree. Additionally, the cation-cation interactions and the positions of all peaks have been accurately recovered, but their intensities show some deviations from the ground truth. The difference between the second and third cation-cation peaks further decrease at higher temperatures.

\begin{figure}[t!]
    \centering
    \includegraphics[width=.49\textwidth]{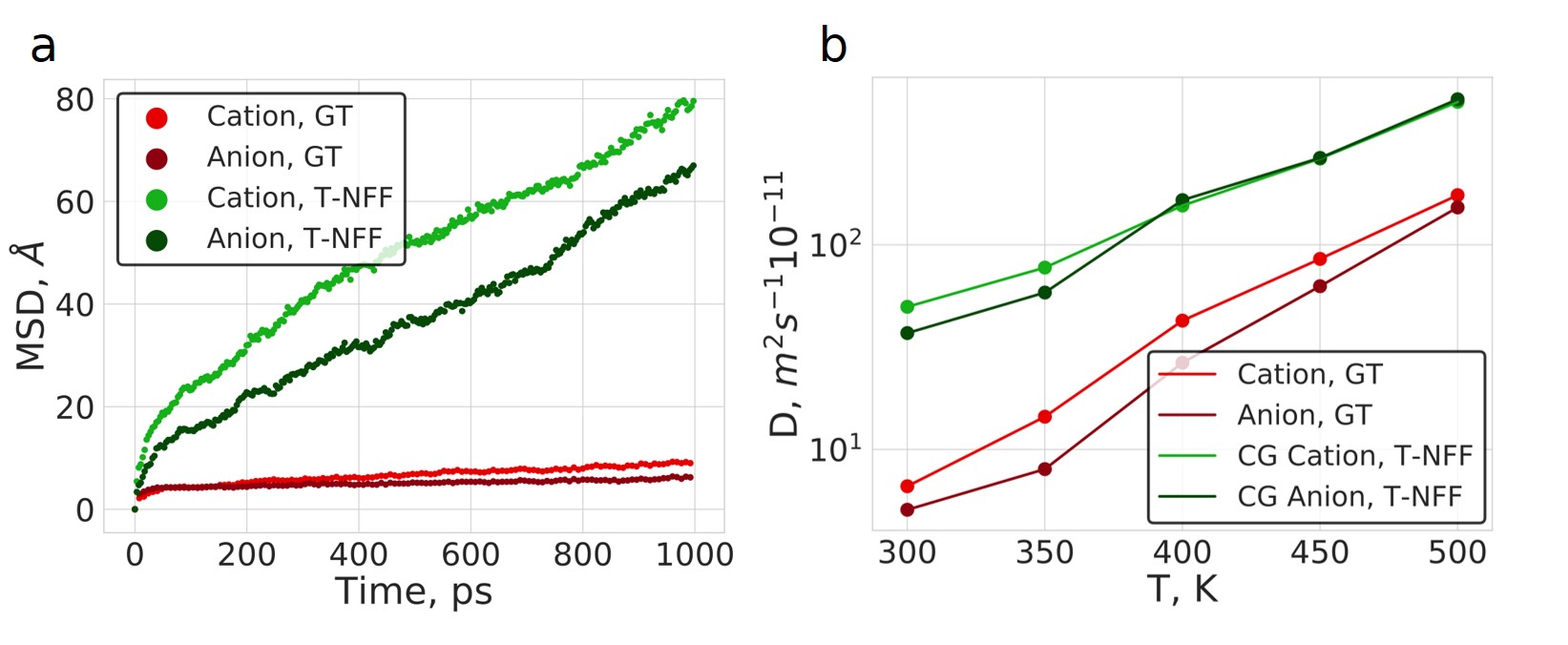}
    \caption{Transport properties of the coarse-grained ionic liquid. a) Mean squared displacement (MSD) over 1 ns long range at 300 K. b) Diffusivity across a temperature range from 300 K to 500 K.}
    \label{arrhenius}
\end{figure}{}

The RDF at 400 K shown in Fig. \ref{RDF}b shows that the model produces stable dynamics at temperatures outside the training set, and can recover structural properties of systems with similar accuracy as at reference temperatures. Bond distance distribution comparisons can be found in the Supplementary Information.

Transport properties such as self-diffusivity are very relevant to the use of ILs as ionic conductors, and  effective CG simulations should ideally recover dynamic properties to the best degree possible. Experiments have shown that the self-diffusivity of the 1-butyl-3-methylimidazolium tetrafluoroborate ionic liquid exhibits Arrhenius behavior \cite{Umecky2005, Sangoro2008ElectricalLiquid}. Figure \ref{arrhenius}a shows the mean-squared displacement calculated by the Einstein relation at 300 K:

% \begin{equation}\label{diffu_eq}
%     D_{a,\mathrm{self}} = \frac{1}{2}\lim_{\tau\to\infty}\frac{\langle[r_{i,a}(t+\tau)-r_{i,a}(t)]^2\rangle}{\tau}.
% \end{equation}

\begin{equation}\label{diffu_eq}
    D_{\mathrm{self}} = \frac{1}{2}\lim_{\tau\to\infty}\frac{\langle[r_{i}(t+\tau)-r_{i}(t)]^2\rangle}{\tau}.
\end{equation}

The evolution of CG T-NFF displacement is about an order of magnitude faster than the ground truth all-atom system. This is because in the CG model, the small, fast movements of hydrogen atoms are averaged out, thus reducing the overall friction of the system and speeding up the dynamics, more discussion change of dynamics in the supplementary information. Fig. \ref{arrhenius}b shows the self-diffusivity across a temperature range from 300 K to 500 K, including out-of-sample simulations at 400 K. Both the ground truth and the T-NFF show similar relations between the cation and the anion. At lower temperatures, the cation has strictly faster dynamics than the anion, but at higher temperatures this difference shrinks. The T-NFF shows the same relation at lower temperatures, but at higher temperatures the difference shrinks faster than for the ground truth. 

\begin{table}[t!]
\begin{tabular}{ccccc}
 & & $E_a$ & (kJ mol\textsuperscript{-1}) \\
Molecule & GT & T-NFF & NFF & CG3 \\
\hline
Cation & 20.7$\pm$1.5  & 14.3$\pm$1.6 & 10.4$\pm$0.6 & 9.4$\pm$0.7   \\ 
Anion  & 21.6$\pm$2.9  & 16.7$\pm$1.8 & 12.2$\pm$0.5 & 8.7$\pm$0.6   \\ 
\end{tabular}
\caption{\label{tbl:energy} Activation energy for diffusivity of the ground truth and CG neural force field for the tested cases. T-NFF and NFF are with the 5 pseudoparticle representation, and CG3 is trained on the T-NFF model with 3 pseudoparticle representation.}
\end{table}

\begin{figure}[t!]
    \centering
    \includegraphics[width=.48\textwidth]{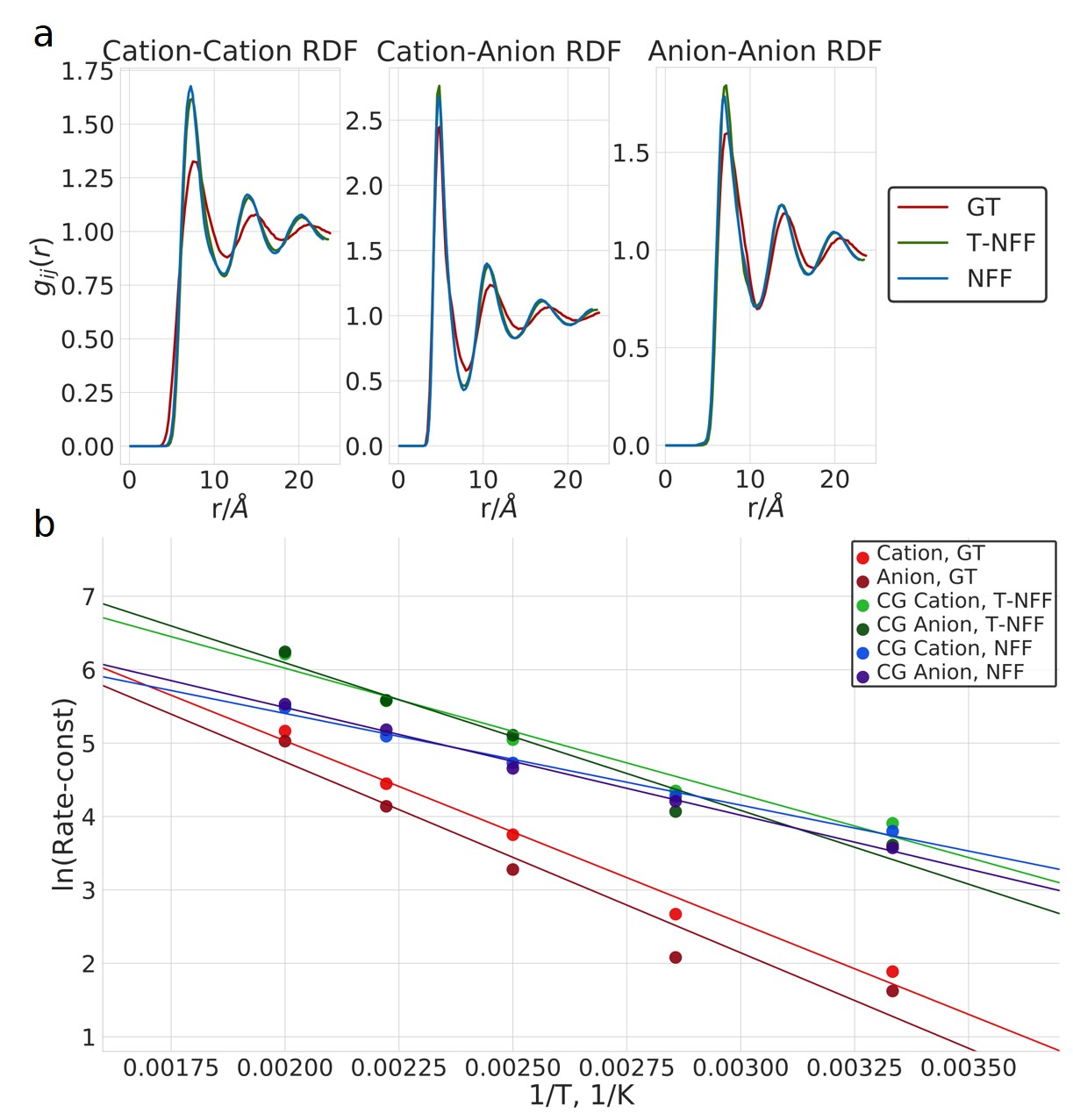}
    \caption{Comparison of a model trained without the temperature embedding layer (NFF) and the temperature transferable model (T-NFF). a) RDF of the ground truth, temperature transferable model and neural force field without the temperature transferability, ran at 400 K. b) Arrhenius plot comparison of the models.}
    \label{compare}
\end{figure}

Figure \ref{compare}b shows the Arrhenius-type relation of the coarse-grained anion and cation. The Pearson correlation coefficient between for the linear fit is -0.981 and -0.982 for cations and anions respectively. The cation and anion curves intersect in the high temperature range as in the ground truth simulation (Table \ref{tbl:energy}), although the intersection occurs at a higher temperature. 

{}

Figure \ref{compare} weighs the impact of the temperature embedding in the model by comparing the outcome of CG simulations from a temperature-independent model (NFF) with the temperature-dependent T-NFF results. Both models were trained on the same data from all-atom simulations at 300, 350, 450 and 500 K and have the same hyperparameters. Figure \ref{compare} shows the RDFs and Arrhenius relations between the ground truth and the models. The two models reproduce the structural properties to a similar degree and the Arrhenius relation of the system is similarly linear. However, as seen from the slopes in Fig. \ref{compare}b and activation energies in Table \ref{tbl:energy}, the temperature-averaged NFF has not learned the temperature effect on dynamics as well as its counterpart and produces a much lower activation energy in addition to accelerated dynamics. It also underestimates the temperature at which cation and anion diffusivities cross.

{}

Figure \ref{compare_3cg} compares the performance the CG3 model (2-bead cation) to the CG5 model. In Fig. \ref{compare_3cg}a, it can be seen that both the 3 and 5 bead models recover structural properties at a temperature that has not been included in the training data. Nevertheless, significant differences are seen in the dynamical properties derived from the Arrhenius relation. Both the models have recovered the linear relation, including the 400K out-of-sample trajectory, but the CG3 model grossly underestimates the activation energy (Table \ref{tbl:energy}). 

\section{Discussion}

We have introduced a novel extension of graph convolutional neural networks to learn many body PMFs by separating short-range intramolecular and longer-range many-body intermolecular terms, and incorporated a temperature transferable embedding to learn the temperature dependence of a many-body PMF. The model has been trained on all-atom simulations of a ionic liquid system, 1-butyl-3-methylimidazolium tetrafluoroborate, whose PMF is known to be difficult to construct. The reconstruction of the structural properties has been achieved to a high degree. Furthermore, the Arrhenius relation for self-diffusivity  has been obtained with the neural force field, but with faster dynamics. The fast dynamics arise due to the averaging of fast motions in the system, which in turn decrease the overall friction of the system. If dynamics cannot be reconstructed exactly, it is preferable that the activation energy for diffusion remain constant between all-atom to CG simulations, and a simple scaling of the dynamics is observed. Other strategies based on spectral matching \cite{Nuske2019} or differentiable simulations \cite{Wang2020b} could be utilized to further condition the encoding function and PMF to reproduce system dynamics.

Results from a single graph convolutional neural network with a single interaction block for both inter- and intra-molecular interaction have not been included in this work for comparison with the novel dual graph convolutional neural network. This is because no stable simulations could be performed with the simpler model. No choice of hyperparameters led to a model that could run coarse-grained IL molecular dynamics simulations without the system collapsing on itself. This is likely because in a coarse-grained model the distances between the beads of a single molecule are on a much larger scale than those of  atoms in the same molecule. These intramolecular distances are then closer to intermolecular distances, and as such it becomes harder for a single interaction layer to separate them and learn the structural and dynamics properties of the system. This highlights the key contribution of the dual graph convolutions introduced in this work. 

A comparison in performance was made between the temperature dependent neural force field (T-NFF) and the neural force field without the temperature learning embedding layer (NFF) (see Fig. \ref{compare}). Both models overestimated the dynamics at lower temperatures equally, but at higher temperatures NFF was significantly slower. The T-NFF showed more homogeneous acceleration of dynamics across the temperature range and as such reproduced the Arrhenius-like temperature dependence much better, largely scaling the overall kinetics by a constant factor of 3-6 at all temperatures. Interestingly, the T-NFF model faithfully captured the loss of linearity in the Arrhenius plot at the lowest temperature that is present in the all-atom simulation and may be related to the system getting trapped in a glassy state. Previous works on on coarse-grained ILs, such as Roy \textit{et al.} \cite{Roy2010} and Moradzadeh \textit{et al.} \cite{Moradzadeh2018}, heavily underestimated dynamics. The model of Roy \textit{et al.} was improved by Merlet \textit{et al.} to attain an agreement of diffusivity \cite{Merlet2012}, although at the cost of accuracy of structural properties. 

Different cation representations were also compared. Both 3- and 5-bead models recovered the structural properties to a similar degree, but the 5 bead model led to better learning of the dynamic temperature relation, despite the overestimation of the rate of the self-diffusion in both cases. 
The model described in this work can accurately learn the many-body PMF of a CG model for a complicated system, is accurate and transferable over a range of temperatures,  requires little user input and can to be applied to other condensed-phase systems, leading to easier, more accurate, and more generalizable CG simulations of complex condensed-phase systems.

\section*{Supplementary Information}
Includes description of training details and hyperparameter selection, further description of parameters of Eq. \ref{eq:loss}, bond distance distribution comparisons for the 5 pseudoparticle model, RDF and bond distance distribution comparisons of the 3 pseudoparticle model. Finally, mean squared displacement comparisons by shifting timescale or temperature are also showed.

\begin{acknowledgments}
The authors thank the MIT Energy Initiative (MITEI), Toyota Research Institute, MIT-IBM and DARPA (Award HR00111920025) for financial support The computations in this paper were executed at the Massachusetts Green High-Performance Computing Center with support from MIT Research Computing..\end{acknowledgments}

\section*{Credit}

This article may be downloaded for personal use only. Any other use requires prior permission of the author and AIP Publishing. This article appeared in the AIP Journal of Chemical Physics and may be found at https://doi.org/10.1063/5.0022431.

\section*{Data and code availablity}
The computer code use and training all-atom trajectories used to train the models are available at \texttt{https://github.com/learningmatter-mit/T-NFF}

\section*{References}

\bibliographystyle{unsrt}  
\bibliography{references}  %%% Remove comment to use the external .bib file (using bibtex).

\clearpage

\pagenumbering{arabic}
% \onecolumn

\section*{Supplementary information}

\subsection*{Training details}
\begin{figure}[h]
    \centering
    \includegraphics[width=.45\textwidth]{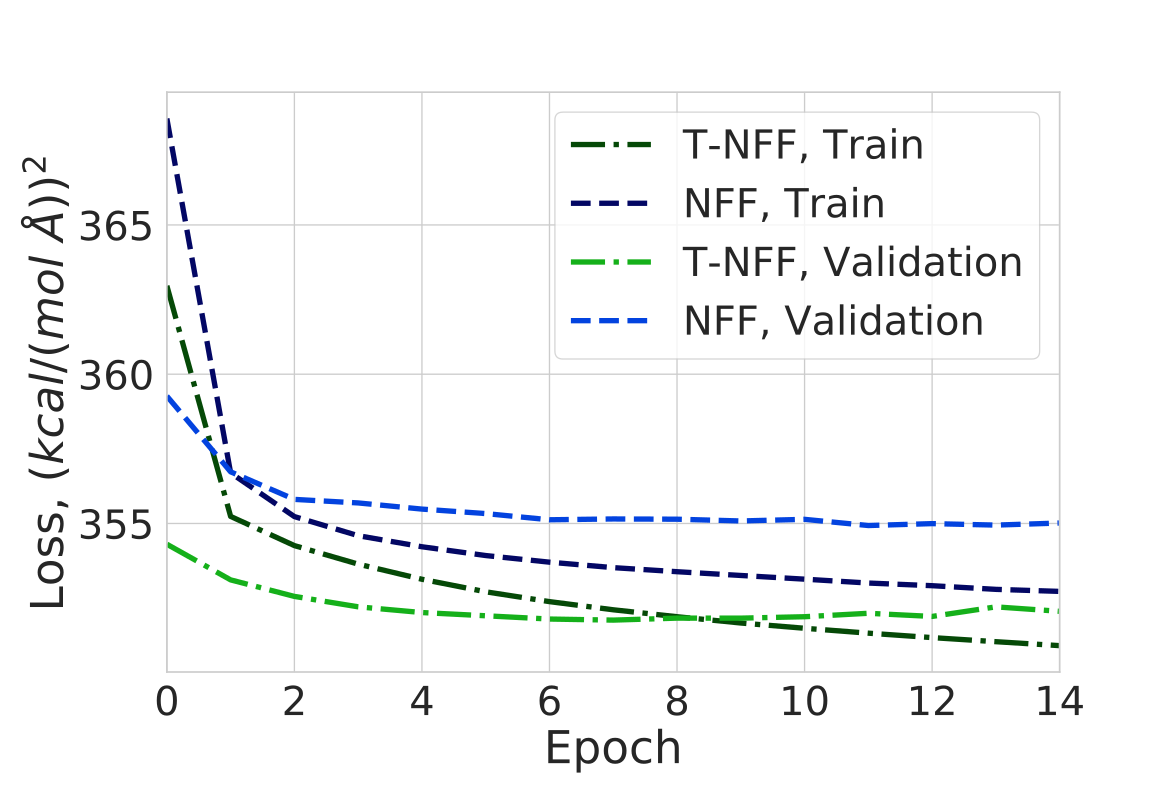}
    \caption{Training and test loss for the model with the temperature transferable embedding (T-NFF) and the one without (NFF)}
    \label{fig:loss}
\end{figure}

Figure \ref{fig:loss} shows the train and test validation for the two models reported in the main text of the paper (T-NFF and NFF). Hyperparameter optimization was done via SigOpt - Bayesian hyperparameter optimization tool.\cite{pmlr-v84-martinez-cantin18a}
While optimizing the hyperparameters on SigOpt, the optimization goal was the MAE of the RDF from the NFF model to the ground truth. An overview of this optimization routine is the following: the hyperparameters are selected, the model is trained and a 100 ps-long MD simulation with the NFF model is run. Then, the MAE between the RDF from this simulation is compared against the ground truth and new hyperparameters are selected based on this criteria.
In the scope of the neural network training, the hyperparameters (with their respective values in brackets) that were optimized include: $d_{inter}$(8) and $d_{intra}$(7) as the neighbor-list distance cutoffs, $n_{atombasis}$(240), $n_{filters}$(256) and $n_{gaussians}$(80) as convolution layer parameters, $n_{interconv}$(2) and $n_{intraconv}$(3) as the number of convolutions for the inter- and intramolecular interaction blocks, $\sigma_{ex}$(7.730579) and $n_{ex}$(7) as the excluded volume parameters and $k$(11.286249) as the bond prior hyperparameter.   
Selected hyperparameters and the trained model for T-NFF can be found on the \texttt{https://github.com/learningmatter-mit/T-NFF} repository.

The selection of the weight parameter $\rho$ from equation 1 (main text), is based on the reconstruction loss and the performance of the CG model according to structural metrics. Generally, low $\rho$ implies high importance given to atomistic reconstruction, which will favor high-resolution CG models with many beads so loss of structural information is minimized. High  $\rho$ implies high importance given to minimizing instantaneous fluctuations in the CG forces and will favor low resolution models with few CG beads which lose more structural information but result in a smoother PMF.  

\begin{figure}[h]
    \centering
    \includegraphics[width=.45\textwidth]{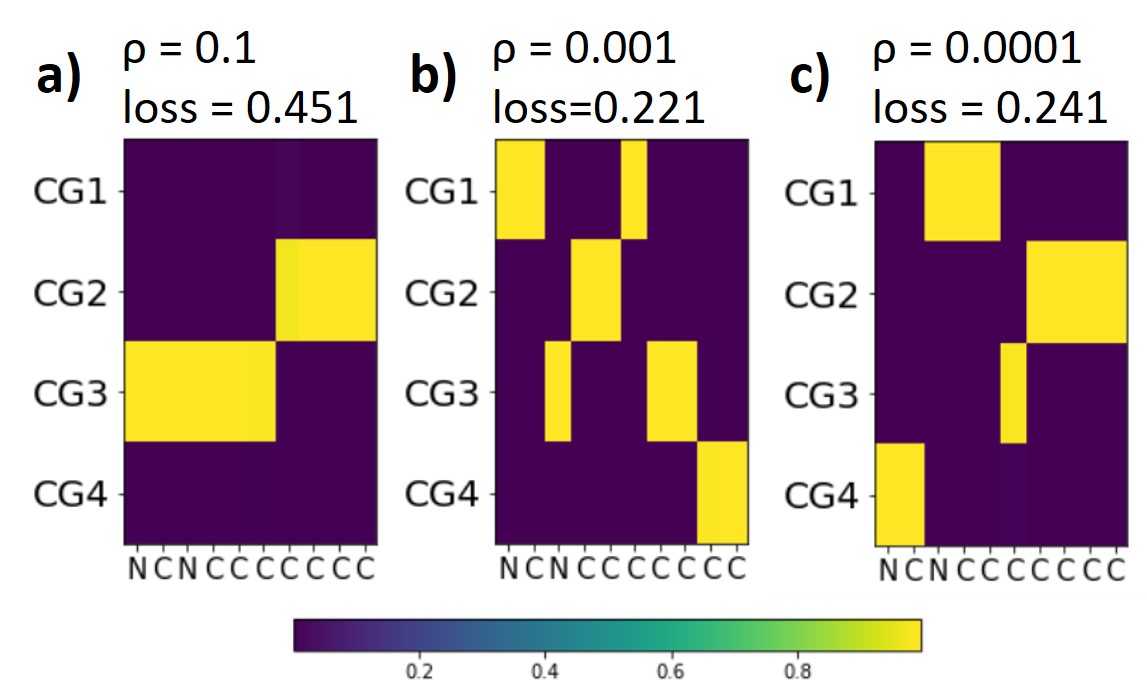}
    \caption{The learned coarse-grained assignments for different $\rho$ parameters. Colors indicate the relative contribution of each atom to the coarse-grained particle.}
    \label{fig:rho}
\end{figure}

In the case of 1-butyl-3-methylimidazolium tetrafluoroborate, at high $\rho$ values, the coarse-grained mapping converges to a mapping with 2 beads, even if the model was created with 4 beads (Figure \ref{fig:rho} a). The minimization of the mean force fluctuation does not take into account the number of mapping dimensions and as such converges to the mapping with the smallest mean forces, ignoring the excess dimensions that had been made available to the model. Having fewer coarse-grained beads to be mapped to leads to more atoms mapped to the same bead, thus averaging out the forces in different directions and as such leading to a smaller mean force value. When the parameter $\rho$ is smaller, the model converges to mappings with larger reconstruction loss (loss .241 (Figure \ref{fig:rho} c) compared to that of 0.221 of the Figure \ref{fig:rho} b).

\subsection*{Simulation details and results}

\begin{figure*}[bth]    
    \centering
    \includegraphics[width=.85\textwidth]{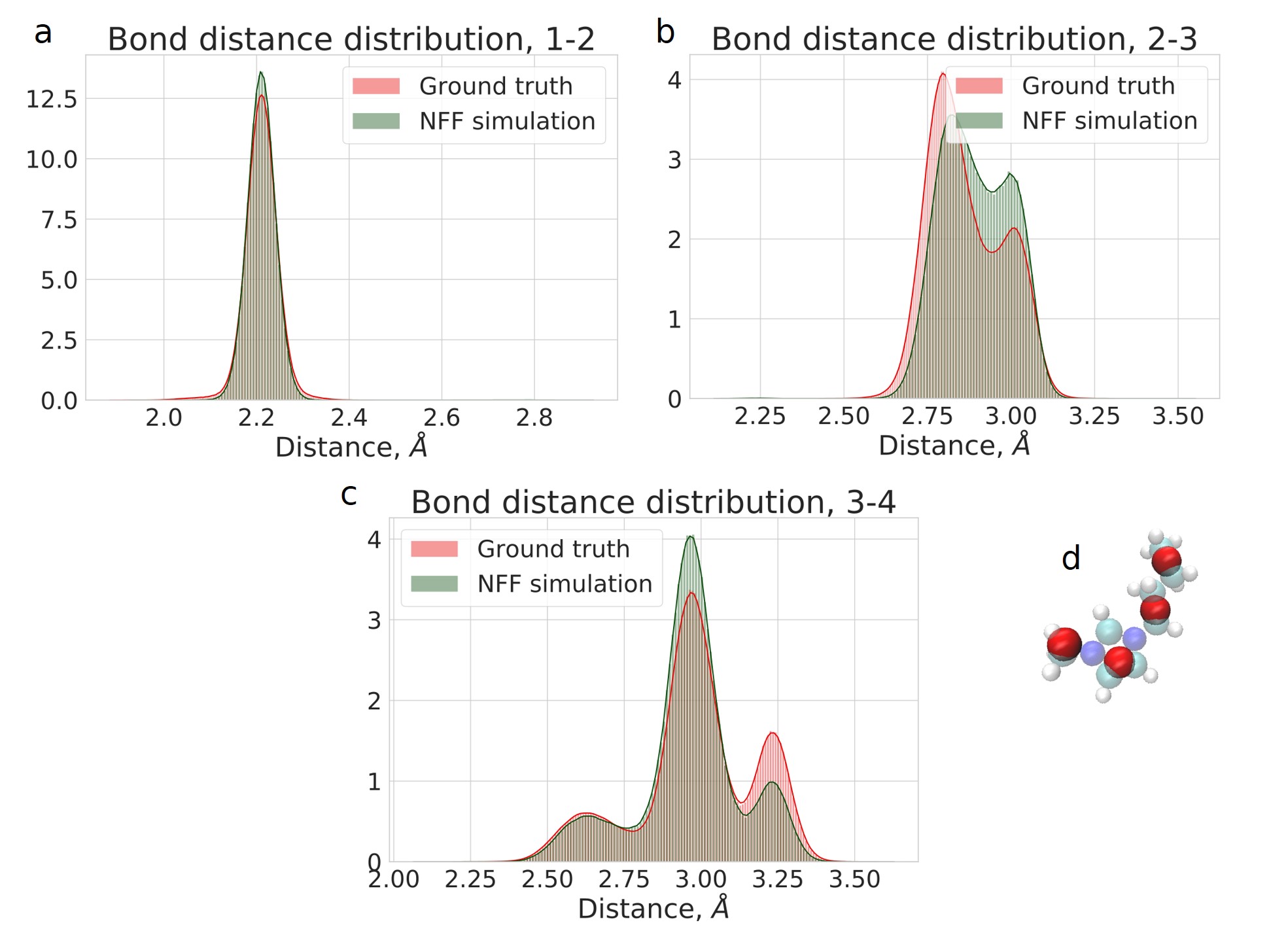}
    \caption{Bond distributions in the coarse-grained cation at 300K. a) The distribution between the first and second pseudoparticle, b) between the second and third pseudoparticle, c) between the third and fourth pseudoparticle. d) the overlay of the auto-encoder generated coarse-grained mapping and positions of pseudoparticles on top of 1-butyl-3-methylimidazolium.}
    \label{bonds}
\end{figure*}

Figure \ref{bonds} shows the learned bond distributions of the temperature transferable graph convolutional neural network at 300 K. The model has properly learned the complex multivariate distributions of the interatomic distances to a high degree.

{}

\begin{figure*}[bth]
    \centering
    \includegraphics[width=.85\textwidth]{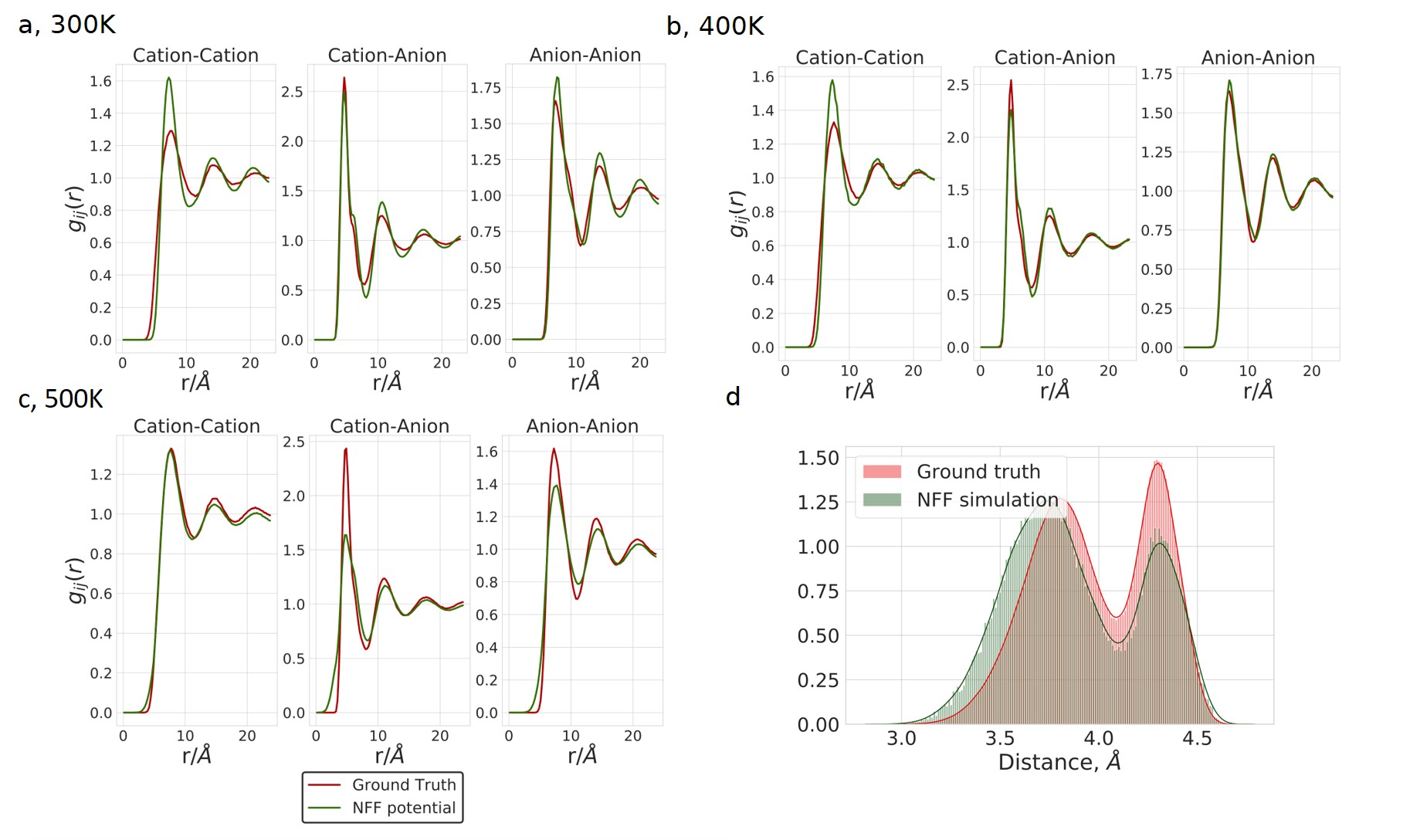}
    \caption{Radial distribution functions and bond distributions of the ionic liquid system with 300 cation and anion molecules, where the cation is represented with 2 pseudoparticles. a) At 300 K, b) at 400 K, and c) at 500K. d) Recovered bond distribution at 300 K of the 2 bead cation representation.}
    \label{rdf_cg3}
\end{figure*}

Figure \ref{rdf_cg3} shows the RDFs across a temperature range for the 2 pseudoparticle representation of the cation as well as the recovered bond distribution. It can be seen from Fig. \ref{rdf_cg3}b that the model has performed reasonably well in recovering the structural properties at 400 K even when this temperature has not been included in the training data.  

{}

\begin{figure*}[h]
    \centering
    \includegraphics[width=.95\textwidth]{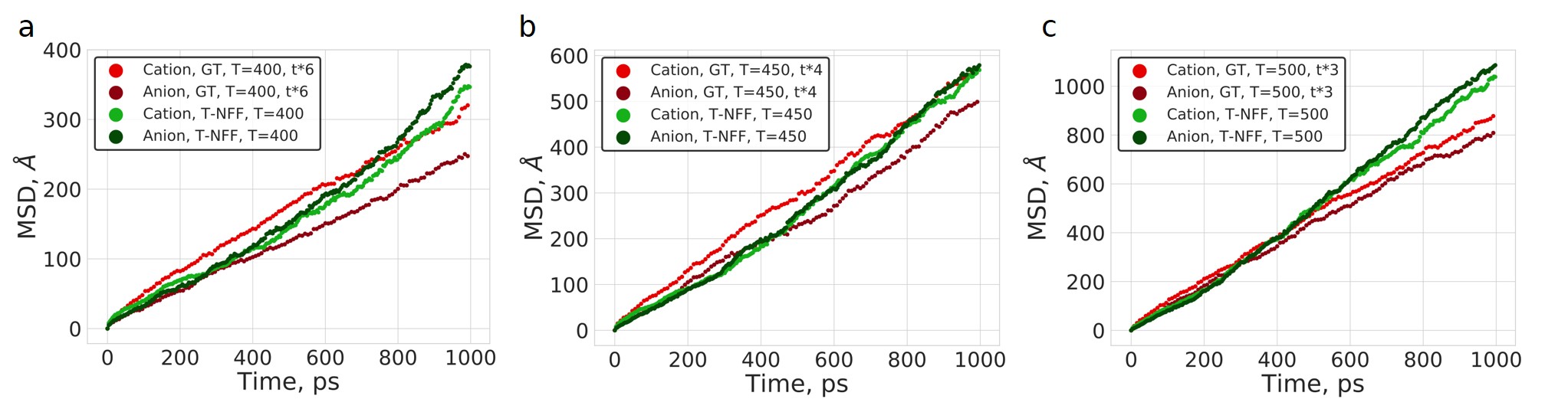}
    \caption{Mean squared displacement comparison of the T-NFF model and the ground truth at the same temperature, but timescale has been shifted by various degrees to have dynamics alignment. a) Both at 400K temperature, but the ground truth time has been scaled by a factor of 6, b) both at 450K temperature, but the ground truth time has been scaled by a factor of 4 and c) both at 500K temperature, but the ground truth time has been scaled by a factor of 3}
    \label{fig:time}
\end{figure*}

Figure \ref{fig:time} shows the mean squared displacement compared of the T-NFF and the ground truth with different scaling factors applied to the ground truth data to obtain overlap of the data. Across the three temperatures (400, 450, 500) the scaling factors vary between 6 and 3. The scaling factor is not exactly constant across the temperatures and ranges for 6 to 3, as the relation between temperature with dynamic properties is not linear. Previous research has outlined a reasoning that coarse-grained simulations generally have some scaling factor regarding the time scale of the simulation. For example in the MARTINI force field, among other works on lipids there are reported scaling factors in the margins from 2-10 from the atomistic simulations,\cite{marrink2004coarse, baron2006configurational, marrink2007martini} in keeping with the scaling found here. Given the value of the activation energy of self-diffusion the acceleration factor is somewhat equivalent to increasing the temperature by 100K. 

\begin{figure*}[bht]
    \centering
    \includegraphics[width=.95\textwidth]{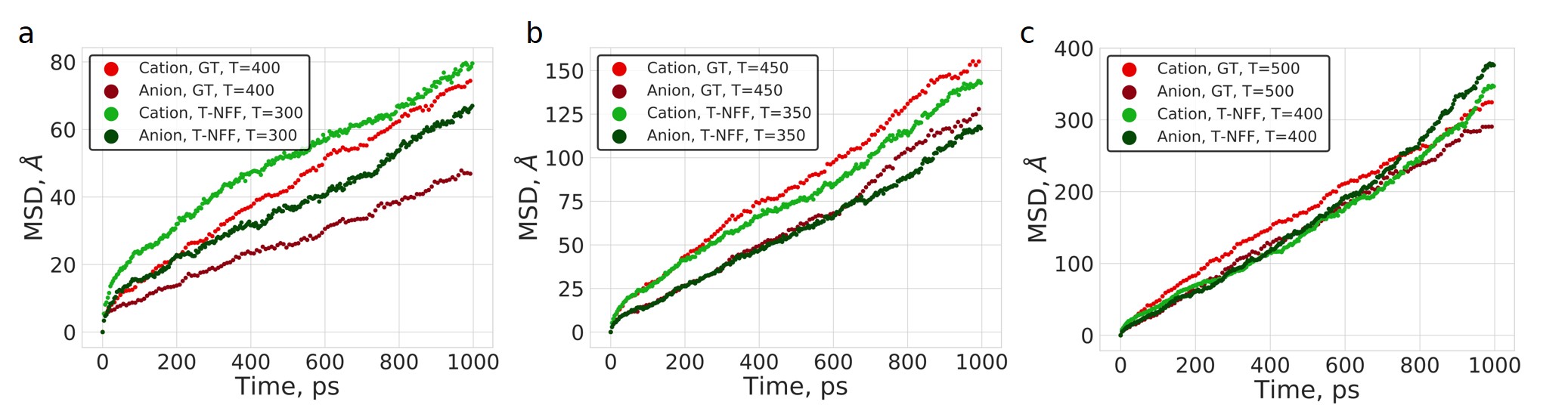}
    \caption{Mean squared displacement comparison of the T-NFF model and the ground truth at 100K higher than the T-NFF simulation. a) T-NFF at 300K and ground truth at 400K, b) T-NFF at 350K and ground truth at 450K and 3) T-NFF at 400K and ground truth at 500K}
    \label{fig:temp}
\end{figure*}

On the other hand, previous ionic liquid research mention the time-scale shift in the coarse-grained models as a temperature shift \cite{Roy2010}. In figure \ref{fig:temp} this is investigated by comparing the T-NFF and the ground truth at a 100K increase. It can be noted that in the temperature range after the 100K increase there is a noticeable agreement between the ground truth and the model. 

\end{document}